\documentclass{article}

\usepackage{arxiv}

\usepackage[utf8]{inputenc} 
\usepackage[T1]{fontenc}    
\usepackage{hyperref}       
\usepackage{url}            
\usepackage{booktabs}       
\usepackage{amsfonts}       
\usepackage{nicefrac}       
\usepackage{microtype}      
\usepackage{lipsum}		
\usepackage{graphicx}
\usepackage{natbib}
\usepackage{doi}
\usepackage{amsthm}
\usepackage{amsmath}
\usepackage{multirow}
\usepackage{makecell}
\usepackage{float}
\usepackage{latexsym}
\usepackage{enumitem}
\usepackage{color}
\usepackage{tikz}
\usepackage{pgfplots}
\usepackage{pgf-umlsd}
\usetikzlibrary{arrows.meta,positioning,matrix,fit,shapes.geometric,intersections,calc,backgrounds}
\pgfplotsset{compat=1.18}

\title{CRiDiT: Instantiating a Run-Time Testbed for Trust Calibration in AI-Infused Systems}

\author{\href{https://orcid.org/0009-0005-2240-721X}{Yuntian~DING} \\
	Centre de Recherche en Informatique\\
    Univ. Paris 1 Panthéon-Sorbonne\\
    Paris, France \\
	\texttt{yuntian.ding@univ-paris1.fr} \\
	\And
	\href{https://orcid.org/0000-0003-1540-2099}{Nicolas~Herbaut} \\
	University of Bordeaux, LaBRI, UMR 5800, F-33400\\
    Talence, France\\
	Centre de Recherche en Informatique\\
    Univ. Paris 1 Panthéon-Sorbonne\\
    Paris, France \\
	\texttt{nicolas.herbaut@u-bordeaux.fr} \\
	\And
	\href{https://orcid.org/0000-0002-1957-0519}{Camille~Salinesi} \\
	Centre de Recherche en Informatique\\
    Univ. Paris 1 Panthéon-Sorbonne\\
    Paris, France \\
	\texttt{camille.salinesi@univ-paris1.fr} \\
}

\renewcommand{\shorttitle}{CRiDiT: Instantiating a Run-Time Testbed for Trust Calibration in AI-Infused Systems}

\hypersetup{
pdftitle={CRiDiT: Instantiating a Run-Time Testbed for Trust Calibration in AI-Infused Systems},
pdfsubject={cs.AI, cs.HC},
pdfauthor={Yuntian~DING, Nicolas~Herbaut, Camille~Salinesi},
pdfkeywords={Trust, Trust Calibration, Trustworthy AI, Computational Trust, AI-Infused System, Design Science Research},
}

\begin{document}
\maketitle

\begin{abstract}
	The integration of AI into larger technical infrastructures has made the alignment of human trust with system trustworthiness, known as trust calibration, a critical engineering concern, since misplaced trust in either direction leads to operational and safety risks. While conceptual frameworks provide a strong foundation for understanding trust calibration, their translation into running systems remains a challenge, because there are few testbeds in which human trust inputs, machine trustworthiness evidence, gap detection and remediation operate together within a closed loop. This paper instantiates CRiDiT (Computational Risk-Sensitive biDirectional Trust) as a run-time testbed, operationalising machine-side trust with Dempster-Shafer Theory and PCR5 redistribution, human-side trust with Subjective Logic, and calibration with a threshold-based trust gap. Following the Design Science Research methodology, we exercise the artifact across three high-stakes scenarios (hiring, financial, legal), producing 144 logged interaction steps across fifteen sessions. The analysis shows that the artifact captures trust calibration dynamics as intended, and reveals three points at which the instantiated policy departs from its design requirements: the machine-side estimate begins from a global benchmark rather than task-relevant evidence; risk-sensitive thresholds do not produce risk-sensitive triggering; and the calibration policy assigns explanatory prompts to over-trust, where corrections narrowed the gap in all 6 observed cases. Since the first two arise from the same design decision, to make the difference of two estimated scalars the calibration criterion, they point toward a common requirement: that the criterion should operate on the evidence rather than on scalars derived from it. The third concerns what follows detection, and shows that the action vocabulary inherited from trust repair does not align with what the interaction logs show to be effective. The work contributes the artifact, a characterisation of its run-time behaviour, and the requirements this characterisation elicits.
\end{abstract}


\section{Introduction}
Artificial intelligence systems are increasingly integrated into human life, creating complex human-AI interactions, from collaborative tools to autonomous agents. Their effectiveness, reliability, and ethical use depend not only on algorithmic accuracy but also on a complex and subjective human factor: trust, understood as a human's belief regarding a system's capacity to perform assigned tasks, together with a willingness to accept uncertainty and vulnerability~\cite{lee2004trust}. As AI is integrated into larger technical infrastructures, what we refer to as an ``AI-infused system''~\cite{amershi2019guidelines}, uncertainty about system behaviour and output has become a major user concern, particularly in high-stakes scenarios where failure can cause significant harm.

We synthesise the core problem as ``misplaced trust causing misassigned power''. Over-trust in a flawed system leads to misuse, while under-trust leaves a capable tool underused~\cite{parasuraman1997humans}. Consider a lawyer using an AI system to analyse a de facto employment relationship dispute and classify the importance of supporting evidence against statutes and previous cases. System trustworthiness here depends on multiple trust factors, which may carry uncertainty and may conflict: privacy protection can limit access to external sources, producing incomplete or inaccurate citations. Since both human perception and system trustworthiness evolve during the interaction, trust calibration is not a property to be established once but a process requiring continuous monitoring.

Conceptual frameworks describe this process well, but translating them into running systems is a different problem, and one that remains largely unaddressed for want of testbeds in which human trust inputs, machine trustworthiness evidence, calibration-state detection, and remediation operate together. We therefore use Design Science Research~\cite{wieringa2014design} as an artifact-centred methodology. Rather than testing causally whether a particular intervention improves trust, we construct a running CRiDiT artifact, evaluate its behaviour in use, and derive requirements for the next design cycle from the gaps exposed by that evaluation. We analyse the interaction logs produced by the artifact through two research questions:
\begin{description}
  \item[\textbf{RQ1:}] What trust calibration dynamics does the artifact capture at run time, and what does its instrumentation leave unresolved?
  \item[\textbf{RQ2:}] Where does the instantiated calibration policy deviate from the design requirements it is intended to satisfy?
\end{description}
\textbf{RQ1} concerns what the testbed observes, and \textbf{RQ2} concerns how faithfully the running artifact realises its own design. 

This paper reports one complete cycle of the design process. Based on the design goals, requirements R1--R6 (Table~\ref{tab:goals_requirements}) are derived and a conceptual model is implemented. Once the instantiated system has been run, a further set of requirements is elicited from its deviations from R1--R6. This second set constitutes the output of the cycle: design knowledge that can only be obtained through construction, and which serves as the input to the next round of investigation. The paper makes three contributions. First, it presents a running instance of CRiDiT that closes the loop between trust input, trust computation, gap detection and remediation (Section~\ref{sec:cridit}). Second, drawing on 144 recorded interaction steps, it characterises the artifact's run-time behaviour, quantifying how the two estimates respond to degradation and to remediation, and on what timescales (Sections~\ref{sec:traced_calibration} through~\ref{sec:remediation}). Third, it evaluates the instantiation against R1--R6, identifies three deviations between the running policy and its design requirements, and translates these together with the requirement-by-requirement assessment into five new requirements for the next design cycle. Each deviation is a property of the design that the conceptual model does not determine and that only a running instantiation makes visible. The analysis is reported in Section~\ref{sec:result}, the implications in Section~\ref{sec:future}, and the conclusions in Section~\ref{sec:conclusion}.

\section{Background}\label{sec:bg}
Trustworthy AI has become a research focus as AI systems are integrated into broader and safety-critical domains, and prior literature has mapped the field through reviews and taxonomies of trustworthy AI and AI-related risks~\cite{afroogh2024trust}, covering regulatory, technical, social and ethical dimensions. Such structural perspectives are valuable for defining trust and categorising its factors, but they offer little purchase on measuring it, and therefore little on aligning human perception with system reliability. The concept of computational trust, introduced by Marsh~\cite{marsh1994formalising}, addresses this by moving from structural definitions toward operational evaluation, providing a way to quantify trust and to integrate and update trust factors over time.

In parallel, prior work on computational trust has proposed mathematical models for representing trust either as a subjective psychological state using probabilistic methods, or through aggregation of weighted trust factors and contextual impacts, or through information fusion techniques for handling trust-related evidence. These methods offer mechanisms to quantify trust and support dynamic updating~\cite{blasch2014trust,fukuchi2023dynamic,roeder2023quantum,hu2024dynamic}. Behavioural evaluation methods emphasise experimentally grounded measures of trust level through behavioural metrics, such as BLOCKIES~\cite{johnson2025higher}. Together, these lines of work provide prerequisites for trust calibration in interactive AI systems. 

Recent research has proposed several methods for trust calibration. Learning-based calibration mechanisms anchor calibration decisions in observed system outcomes, for example, Yu et al.~\cite{yu2017user} illustrate how trust evolves in relation to system performance. Predictive models consider human trust as a temporal process and aim to anticipate future human behaviour in order to identify potential miscalibration during human-AI interaction, such as Wang et al.'s work~\cite{wang2022will}. Other frameworks focus on the role of explainability and confidence cues in influencing human trust, including AXTF (Adaptive Explainability Trust Framework)~\cite{fernando2025adaptive}, and Yu et al.'s computational framework~\cite{yu2025enhancing}. These methods typically address individual dimensions of trust calibration, such as explainability, performance evaluation, or trust prediction in isolation. By contrast, conceptual calibration frameworks such as TCMM (Trust Calibration Maturity Model)~\cite{steinmetz2025trust}, IMPACTS (Intention, Measurability, Performance, Adaptivity, Communication, Transparency, Security)~\cite{hou2025impacts} and CRiDiT (Computational Risk-Sensitive biDirectional Trust) structure trust factors, risks, and (re)calibration processes at a system level, providing guidance for analysis and evaluation. 

Translating these frameworks into a running AI-infused system introduces constraints they do not address. Trust evaluation must account for contextual risk, and for the fact that trust changes during interaction rather than being assigned once. The two sides of the relationship must be represented in a form that permits comparison, since a calibration decision rests on that comparison, and the decision must resolve to a concrete remediation action.

This paper extends the CRiDiT framework introduced in our systematic literature review~\cite{ding2026toward} by implementing its layered architecture in a running chatbot prototype. CRiDiT provides the operational basis for this work because it explicitly connects trust inputs, computational trust functions, and system-embedded remediation actions, thereby enabling instantiation. A parallel line of work addresses trust repair, that is, the restoration of trust after a system failure. De Visser et al.~\cite{de2018automation} identify explanation and correction as its core strategies, which are also the actions most calibration designs inherit when they come to specify what the system should actually do.

\section{The CRiDiT Artifact}\label{sec:cridit}
CRiDiT is designed to satisfy three goals: risk sensitivity, so that calibration responds to the stakes of the context; dynamic trust management, so that both estimates update during interaction; and remediation actions, so that detected miscalibration produces a concrete system response. Table~\ref{tab:goals_requirements} states the requirements that operationalise each goal, derived from these goals before the artifact was built.

\begin{table}[h]
  \caption{Design goals and requirements. R1--R6 serve as implementation criteria and as the basis against which the instantiation is assessed in Section~\ref{sec:requirements}.}
  \label{tab:goals_requirements}
  \begin{tabular}{p{0.13\columnwidth} p{0.03\columnwidth} p{0.76\columnwidth}}
    \toprule
    \textbf{Design Goal} & \multicolumn{2}{l}{\textbf{Requirements}} \\
    \midrule
    \multirow{2}{*}{Risk Sensitivity} & R1 & CRiDiT shall collect perceived risk; it directly influences human trust computation. \\
    & R2 & CRiDiT shall trigger remediation actions based on contextual risk level, with higher risk demanding faster response. \\
    \addlinespace
    \multirow{3}{*}{\makecell[tl]{Dynamic Trust\\Management}} & R3 & CRiDiT shall update trust scores and calibration status in real time based on interactions. \\
    & R4 & CRiDiT shall quantify subjective human trust signals in a form that is comparable with machine trust scores. \\
    & R5 & CRiDiT shall quantify and fuse technical evidence to compute the machine trust score, starting from a benchmark performance baseline as a prior in the absence of task-specific evidence. \\
    \addlinespace
    \makecell[tl]{Remediation\\Actions} & R6 & CRiDiT shall map calibration status (over-trust, under-trust, well-calibrated) to concrete remediation actions. \\
    \bottomrule
  \end{tabular}
\end{table}
Section~\ref{sec:future} returns to them with a second set derived from the instantiation itself. 

\subsection{CRiDiT Architecture}
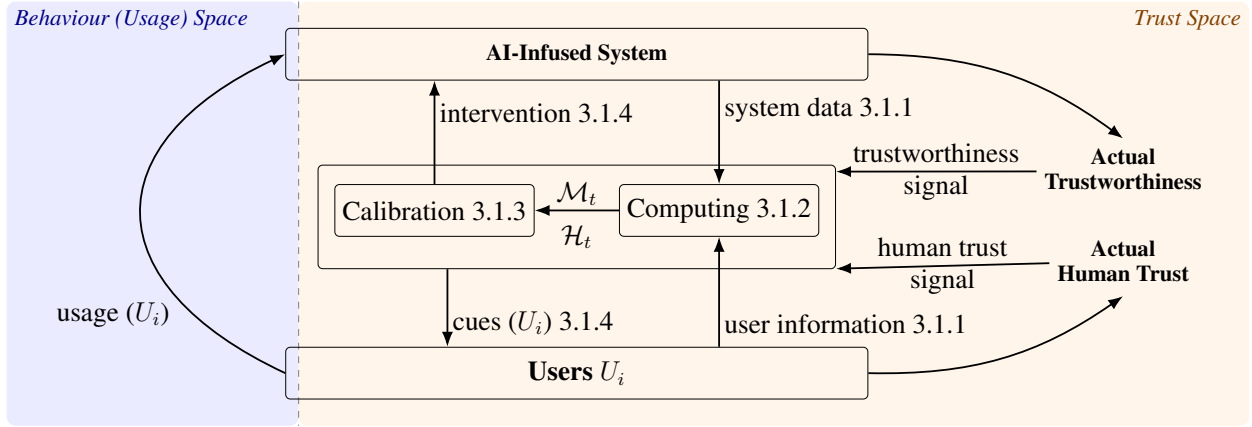
\begin{figure}[!h]
  \centering
  \resizebox{\columnwidth}{!}{
  \begin{tikzpicture}[
    font=\large,
    >=Latex,
    node distance=10mm and 18mm,
    block/.style={draw, rounded corners=2pt, minimum width=34mm, minimum height=12mm, align=center},
    smallblock/.style={draw, rounded corners=2pt, minimum width=18mm, minimum height=8mm, align=center},
    user/.style={circle, draw, minimum size=5mm, inner sep=0pt},
    flow/.style={->, thick, rounded corners},
    feedback/.style={->, thick, rounded corners, bend left=18},
    labelbox/.style={fill=white, inner sep=2pt, align=center}
  ]

  \node[smallblock, minimum width=90mm] (aisystem) {\normalsize \textbf{AI-Infused System}};

  \node[block,  minimum width=80mm, minimum height=16mm, below=13mm of aisystem] (evaluation) {};
  \node[smallblock] (cal) at ($(evaluation.center)+(22mm,1mm)$) {Computing~\ref{subsec:cridit_computing}};
  \node[smallblock] (cali) at ($(evaluation.center)+(-22mm,1mm)$) {Calibration~\ref{subsec:cridit_cali}};
  \node[below=0mm of evaluation.center] {$\mathcal{H}_t$};
  \draw[flow] (cali.north) -- node[right, labelbox, fill=none, yshift=3mm] {intervention~\ref{subsec:cridit_output}} ($(aisystem.south)+(-22mm,0)$);
  \draw[flow] ($(aisystem.south)+(22mm,0)$) -- node[right, labelbox, fill=none, yshift=3.5mm] {system data~\ref{subsec:cridit_input}} (cal.north);
  \draw[<-, thick] (cali.east) -- node[above, labelbox, fill=none] {$\mathcal{M}_t$} (cal.west);

  \node[smallblock, minimum width=90mm, below=12mm of evaluation] (userslabel) {\textbf{Users} $U_i$};
  \draw[flow] ([xshift=22mm] userslabel.north) -- node[right, labelbox, fill=none, yshift=-5mm] {user information~\ref{subsec:cridit_input}} (cal.south);

  \node[right=31mm of evaluation, yshift=20] (trustworthiness) {\normalsize \makecell{\textbf{Actual}\\\textbf{Trustworthiness}}};
  \node[below=4mm of trustworthiness] (humantrust) {\normalsize \makecell{\textbf{Actual}\\\textbf{Human Trust}}};

  \draw[flow] (aisystem.east) to [bend left=18] (trustworthiness.north);
  \draw[feedback] (userslabel.east) to [bend right=18] (humantrust.south);

  \draw[flow] (trustworthiness.west) -- node[midway, labelbox, fill=none] {\makecell{trustworthiness\\signal}} ([yshift=7mm] evaluation.east);
  \draw[flow] (humantrust.west) -- node[midway, labelbox, fill=none] {\makecell{human trust\\signal}} ([yshift=-8mm] evaluation.east);

  \draw[flow] ([xshift=-20mm] evaluation.south) -- node[right, labelbox, pos=0.45, fill=none, yshift=-3mm] {cues ($U_i$)~\ref{subsec:cridit_output}} ([xshift=-20mm] userslabel.north);

  \draw[flow] (userslabel.west) to[out=155, in=205, looseness=1.7] node[left, labelbox, pos=0.2, xshift=-8, fill=none] {usage ($U_i$)} ($(aisystem.west)+(0,0mm)$);

  \begin{pgfonlayer}{background}
      \coordinate (arcbulge) at ($(evaluation.west)+(-42mm,-10mm)$);
      \node[fit=(aisystem)(evaluation)(cal)(cali)(userslabel)(trustworthiness)(humantrust)(arcbulge), inner xsep=6mm, inner ysep=4mm] (bbox) {};
      \coordinate (splitline) at ($(bbox.west)!0.235!(bbox.east)$);
      \fill[blue!8, rounded corners=4pt] (bbox.north west) rectangle (bbox.south -| splitline);
      \fill[orange!8, rounded corners=4pt] (bbox.north -| splitline) rectangle (bbox.south east);
      \draw[dashed, gray] (bbox.north -| splitline) -- (bbox.south -| splitline);
      \node[anchor=north west, font=\normalsize\itshape, text=blue!50!black] at (bbox.north west) {Behaviour (Usage) Space};
      \node[anchor=north east, font=\normalsize\itshape, text=orange!50!black] at (bbox.north east) {Trust Space};
    \end{pgfonlayer}
  \end{tikzpicture}}
  \caption{The interaction closed-loop CRiDiT instantiates. In the behaviour space (left), users interact with the AI-infused system; the system's actual trustworthiness and the user's actual trust are properties of that interaction which CRiDiT never observes directly. In the trust space (right), it approximates both from the signals available to it: system data yields the machine trustworthiness estimate $\mathcal{M}_t$ in the computing layer, user information yields the human trust estimate $\mathcal{H}_t$, and the calibration layer compares them. The loop closes through two channels, cues returned to the user and interventions returned to the system.}\label{fig:cridit_interaction_loop}
\end{figure}

Figure~\ref{fig:cridit_interaction_loop} shows the closed loop CRiDiT instantiates. The loop spans two spaces. In the behaviour space, users interact with the AI-infused system, and each side of that interaction has a property the system never observes directly: the system's actual trustworthiness on the task at hand, and the user's actual trust in it. In the trust space, CRiDiT approximates both from the signals available to it, producing the run-time estimates $\mathcal{M}_t$ and $\mathcal{H}_t$, compares them in the calibration layer, and returns cues to the user and interventions to the system. Calibration therefore operates on estimates of the two quantities rather than on the quantities themselves, and the fidelity of that approximation bounds what the loop can achieve.

Table~\ref{tab:notation} presents the notations used in the following sections to facilitate understanding. 

\begin{table}[h]
  \caption{CRiDiT Notations}\label{tab:notation}
  \begin{tabular}{p{0.25\columnwidth} p{0.7\columnwidth}}
    \toprule
    \textbf{Notations} & \textbf{Description} \\
    \midrule
    $X=\{\text{T}$, $\text{UT}\}$ & The set of all possible states: trustworthy $\text{T}$ or untrustworthy $\text{UT}$. \\
    $2^X=\{\emptyset, \{\text{T}\}, \{\text{UT}\}, \{\text{T}, \text{UT}\}\}$ & The power set that includes all the possible combinations of elements in the set $X$. \\
    $\mathcal{E}^{\mathrm{tech}+}_t$, $\mathcal{E}^{\mathrm{tech}-}_t$,& Positive and negative evidence sets. \\
    $\mathcal{E}^{\mathrm{tech}}_t = \mathcal{E}^{\mathrm{tech}+}_t \cup\mathcal{E}^{\mathrm{tech}-}_t$ & The combined technical evidence set. \\
    $\mathcal{E}_t = \mathcal{E}^{\text{tech}}_t \cup\mathcal{R}_t$ & The union set of technical evidence and perceived risks collected during interaction.\\
    $N_{\mathcal{E}_t}$ & The number of evidence collected (trust factors and perceived risks) at time $t$. \\
    $e_i$, $X_{e_i}$, $m_{e_i}$ & The $i^\text{th}$ evidence in $\mathcal{E}_t$, its state, and its belief mass.\\
    $X_{\mathcal{E}_t}$ & The set of all the $X_{e_i}$. \\
    $m_t^{\text{cons}} (x)$ & The belief mass of state $x$ resulting from consensus fusion. \\
    $m_t^{\text{conf}}$ & The belief mass resulting from conflict fusion.\\
    $m_{\mathcal{E}_t} (x), x\in2^X \setminus\{\emptyset\}$ & The global belief mass on state $x$. \\
    $bel_t(\{T\})$ & The belief of trust state at time $t$. \\
    $pl_t(\{T\})$ & The plausibility of trust state at time $t$. \\
    $\mathcal{M}_t$ & Machine trustworthiness estimate in $[0,1]$. \\ 
    $b_t^T, d_t^T, u_t^T, a^T$ & Belief, disbelief, and uncertainty in $\{T\}$, and base rate, at time $t$. \\
    $\omega^{\text{beh}}_t$, $\omega^{\text{subj}}_t$ & Opinions based on behavioural and subjective inputs. \\
    $\omega_t$ & Fused opinion $(b,d,u,a)$ obtained from behavioural and subjective inputs.\\
    $\mathcal{H}_t$ & Human trust estimate in $[0,1]$.\\
    \bottomrule
  \end{tabular}
\end{table}

\subsubsection{Input Layer}\label{subsec:cridit_input}
The input layer collects trust factors across three dimensions: socio-ethical considerations, technical characteristics, and human inputs~\cite{bach2024systematic}.

\textit{System-Related Trust Factors} are derived from objective evidence that enables cognitive judgement of trustworthiness. Positive evidence includes reliability metrics; negative evidence includes failures and detected flaws; uncertainty evidence covers cases where the origin of a problem is ambiguous, for instance when an incomplete response could reflect a system limitation or an incomplete prompt. These map to belief masses on $\{T\}$, $\{UT\}$ and $\{T,UT\}$ respectively in the DST evidence set $\mathcal{E}_t$.

\textit{Human-Related Trust Factors} encompass both external influences, such as the provider's reputation and prior experience with similar systems, and internal variabilities, such as dispositional traits. Measuring these is challenging because individual subjectivity makes many proxies unreliable (heart rate, for instance, reflects stress but not necessarily trust). In this work we treat adoption of system outputs as the primary behavioural signal and self-reported ratings as the subjective input.

In addition to these trust factors, Lee and See's~\cite{lee2004trust} definition of trust emphasises its strong relationship with risk. Risk analysis must incorporate threats~\cite{yampolskiy2016taxonomy} and their potential negative impacts~\cite{weidinger2022taxonomy}, as well as vulnerabilities at three levels: system-level, model-level, and human-communication-level.

\subsubsection{Computing Layer}\label{subsec:cridit_computing}
This layer computes $\mathcal{M}_t$ from technical evidence and $\mathcal{H}_t$ from human inputs.
\paragraph{Machine-side Trust Evaluation $\mathcal{M}_t$.}
Trust factors can conflict: high accuracy may coexist with privacy risk, and negative evidence may be swamped by positive evidence under Dempster's normalisation rule. The Dempster-Shafer Theory (DST) handles such conflicts through evidence fusion~\cite{blasch2014trust}. For a set of $N_{\mathcal{E}_t}$ evidence items, the consensus mass and conflict mass are obtained by conjunctive combination. To avoid discarding strong negative evidence, we redistribute the conflict using the Proportional Conflict Redistribution rule n°5 (PCR5)~\cite{blasch2014static,smarandache2004proportional,smarandache2006proportional}, as given in Equation~\ref{eq:pcr5}.

Formally, $\forall\ x \in2^X\setminus\{\emptyset\},\;\forall\ X_{\mathcal{E}_t} \subseteq\ 2^X$: 
\begin{equation*}
  m_t^{\text{cons}}(x) = \sum_{\substack{X_{\mathcal{E}_t} \\ \bigcap\limits_{i=1}^{N_{\mathcal{E}_t}} X_{e_i} = x }} \prod_{i=1}^{N_{\mathcal{E}_t}} m_{e_i}(X_{e_i})
\end{equation*}
and the belief mass for conflict conjunctive is expressed as:
\begin{equation*}
  m_t^{\text{conf}} = \sum_{\substack{X_{\mathcal{E}_t} \\ \bigcap\limits_{i=1}^{N_{\mathcal{E}_t}} X_{e_i} = \emptyset }} \prod_{i=1}^{N_{\mathcal{E}_t}} m_{e_i}(X_{e_i})
\end{equation*}
\begin{equation}
  m_{\mathcal{E}_t}(x) = m_t^{\text{cons}}(x) + \sum_{\substack{X_{\mathcal{E}_t} \\
  \bigcap\limits_{i=1}^{N_{\mathcal{E}_t}} X_{e_i} = \emptyset \\
  X_{e_i} = x}}
  \left(\frac{m_{e_i}(x)^2 \prod\limits_{\substack{k=1\\k\neq i}}^{N_{\mathcal{E}_t}}
  m_{e_k}(X_{e_k})}{m_{e_i}(x) +
  \sum\limits_{\substack{k=1\\k\neq i}}^{N_{\mathcal{E}_t}}
  m_{e_k}(X_{e_k})}\right)
  \label{eq:pcr5}
\end{equation}

We derive a scalar estimate via the Pignistic transform~\cite{smets1994transferable}. For the binary frame $X = \{T, UT\}$ this equals the midpoint of the belief-plausibility interval:
\begin{equation*}
  \mathcal{M}_t = \text{BetP}_t(\{T\}) = \frac{bel_t(\{T\}) + pl_t(\{T\})}{2}
\end{equation*}
where $bel_t(\{T\}) = m_{\mathcal{E}_t}(\{T\})$ and $pl_t(\{T\}) = m_{\mathcal{E}_t}(\{T\}) + m_{\mathcal{E}_t}(\{T, UT\})$.

\paragraph{Human Trust Perception $\mathcal{H}_t$.}
We adapt Subjective Logic~\cite{josang2001logic,josang2016subjective} to represent human trust as an opinion $\omega_t = (b_t^T, d_t^T, u_t^T, a^T)$ over the proposition that the system is trustworthy. Behavioural inputs (acceptance or rejection) produce a binomial opinion; subjective inputs (self-reported ratings) are mapped via the qualitative-matrix method~\cite{josang2016subjective}, where likelihood is the average of reliability and predictability ratings, confidence is taken from self-confidence, and the base rate is adjusted by task criticality. The two opinions are fused with weighted consensus using raw weights $w^{\text{beh}} = 0.7$ and $w^{\text{subj}} = 0.3$; after the power transform ($\gamma = 2$) and normalisation the effective weights are $0.845$ and $0.155$. The human trust score is the expected probability of trustworthiness from the fused opinion:
\begin{equation*}
  \mathcal{H}_t = E(\omega_t) = b_t^T + a^T u_t^T
\end{equation*}
The full derivation of the binomial opinion and the fuzzy-logic mapping for $\mathcal{H}_0$ are given in Appendix~\ref{app:human_trust}.

\subsubsection{Trust Calibration}\label{subsec:cridit_cali}
The gap $\Delta_t = \mathcal{H}_t - \mathcal{M}_t$ measures the signed difference between the human trust estimate and the machine trustworthiness estimate. A threshold $\tau$ defines three calibration states:
\begin{itemize}
  \item $|\Delta_t| \leq \tau$: well-calibrated.
  \item $\Delta_t > \tau$: over-trust where the participant trusts the system more than its estimated trustworthiness warrants.
  \item $\Delta_t < -\tau$: under-trust where the participant trusts the system less than its estimated trustworthiness warrants.
\end{itemize}

The threshold is a design parameter: a lower value increases sensitivity to miscalibration and is appropriate for higher-risk contexts.

\subsubsection{Output Layer}\label{subsec:cridit_output}
The output layer communicates two categories of information to the participant: trust indicators ($\mathcal{M}_t$, $\mathcal{H}_t$, $\Delta_t$, and calibration status), and remediation actions triggered when a threshold crossing is detected. Remediation acts through two mechanisms: a static warning displayed to the participant, and a prompt issued to the language model. The calibration state determines which warning and which prompt type are used.

\subsection{Prototype Implementation}
We instantiate the CRiDiT architecture within a chatbot prototype to operationalise R1--R6 across three high-stakes application scenarios: corporate financial, legal, and hiring. The prototype integrates with the OpenAI API under token constraints.

\subsubsection{Operationalisation of CRiDiT Constructs}\label{subsec:operationalisation}
\paragraph{Machine Trustworthiness Estimate and Human Trust Estimate.} The two estimates are driven by different channels, governing both what each can represent and when each is revised.

$\mathcal{M}_t$ is driven by observer annotation. Before a prompt is sent to the model, an observer selects a trust factor from the scenario's pre-flight questionnaire, assigns a polarity and enters a severity $\sigma \in [0,1]$. Each event starts from a polarity-dependent base mass: positive events begin with $(m(\{T\})=0.7,\, m(\{UT\})=0.1,\, m(\{T,UT\})=0.2)$ and negative events with $(m(\{T\})=0.1,\, m(\{UT\})=0.7,\, m(\{T,UT\})=0.2)$. Severity scales the dominant mass by $0.75+0.25\sigma$ and the minor mass by $1-0.25\sigma$, with the remainder assigned to $\{T,UT\}$. A fixed context weight of 0.6 then multiplies $m(\{T\})$ and $m(\{UT\})$ by $0.72$, with the remainder again added to $\{T,UT\}$. Polarity and severity are observer judgements of the intervention's trust relevance rather than measurements of output quality, which bounds what $\mathcal{M}_t$ can be taken to represent.

The label field records the intervention type using four categories: \textit{correction}, \textit{explanation (complementary)} for information supplements that do not alter response correctness, \textit{explanation (uncertainty communication)} for prompts that make the system's limits explicit, and \textit{violation} for injected flaws. Three events carry no logged prompt and their type cannot be independently verified from the text; all other labels were verified against the prompt content and corrected where needed. This has consequences for the three unclassifiable events in the remediation analysis in Section~\ref{sec:remediation}.

The base masses, severity scaling formula, and context weight are heuristic choices made to produce plausible trust dynamics during piloting. They were not derived from empirical data nor independently validated; their sensitivity to variation is not characterised in this study and constitutes an open question for the next design cycle.

$\mathcal{H}_t$ is driven by participant input. The pre-flight questionnaire initialises $\mathcal{H}_0$ as the average of rescaled Likert responses, where each item score $\text{ans} \in \{1,\ldots,7\}$ maps to $(\text{ans}-1)/6$, with fuzzy membership functions assigning a base rate, uncertainty and per-participant threshold. During interaction, each step contributes an adoption decision and four self-reported ratings. The adoption decision follows the binomial model; the ratings are mapped to a Subjective Logic opinion via a custom scalar transformation. The two opinions are fused with weighted consensus, with effective weights of 0.845 and 0.155 after the power transform. The consensus base rate omits the correction term from J{\o}sang's original definition~\cite{josang2016subjective}, so the result is bounded to $[0,1]$ by clipping rather than by construction. The human trust score is $\mathcal{H}_t = b_t + a^T \cdot u_t$. Full derivations are in Appendix~\ref{app:human_trust}.

The two estimates differ in source (observer annotation versus participant report), granularity (factor-level polarity and severity versus a seven-point scale over four dimensions), and timing (before versus after the participant has seen the response). The timing difference has consequences for how calibration states in the interaction logs should be read, and we return to it in Section~\ref{sec:traced_calibration}.

\paragraph{Calibration settings.} Threshold values were established through piloting: $\tau = 0.06$ for legal, $0.07$ for financial, and $0.08$ for hiring. The pre-flight fuzzy membership functions compute a per-participant threshold and base rate, but these were not used during the study; the fixed per-scenario threshold was applied throughout.

The initial machine trust score $\mathcal{M}_0 = 0.925$ is taken from the MMLU score of the underlying model, initialised as $(m(\{T\})=0.85,\, m(\{UT\})=0, m(\{T,UT\})=0.15)$, which recovers 0.925 via the Pignistic transform. Section~\ref{sec:requirements} discusses the consequences of this choice.

\paragraph{Remediation actions.} Static warnings are displayed according to calibration state. For over-trust: \textit{``Please verify the output carefully. The system can make mistakes.''} For under-trust: \textit{``The system performs well in general. You can rely on it to improve efficiency.''} No warning is shown when well-calibrated. In addition, the calibration state determines the prompt type issued to the language model: corrective prompts under under-trust when a flaw is identified, or explanatory prompts when the participant requests additional transparency; explanatory prompts communicating uncertainty and limitations under over-trust. Section~\ref{sec:requirements} assesses how this mapping held in practice.

\subsubsection{Interaction Flow}
\begin{figure}[t]
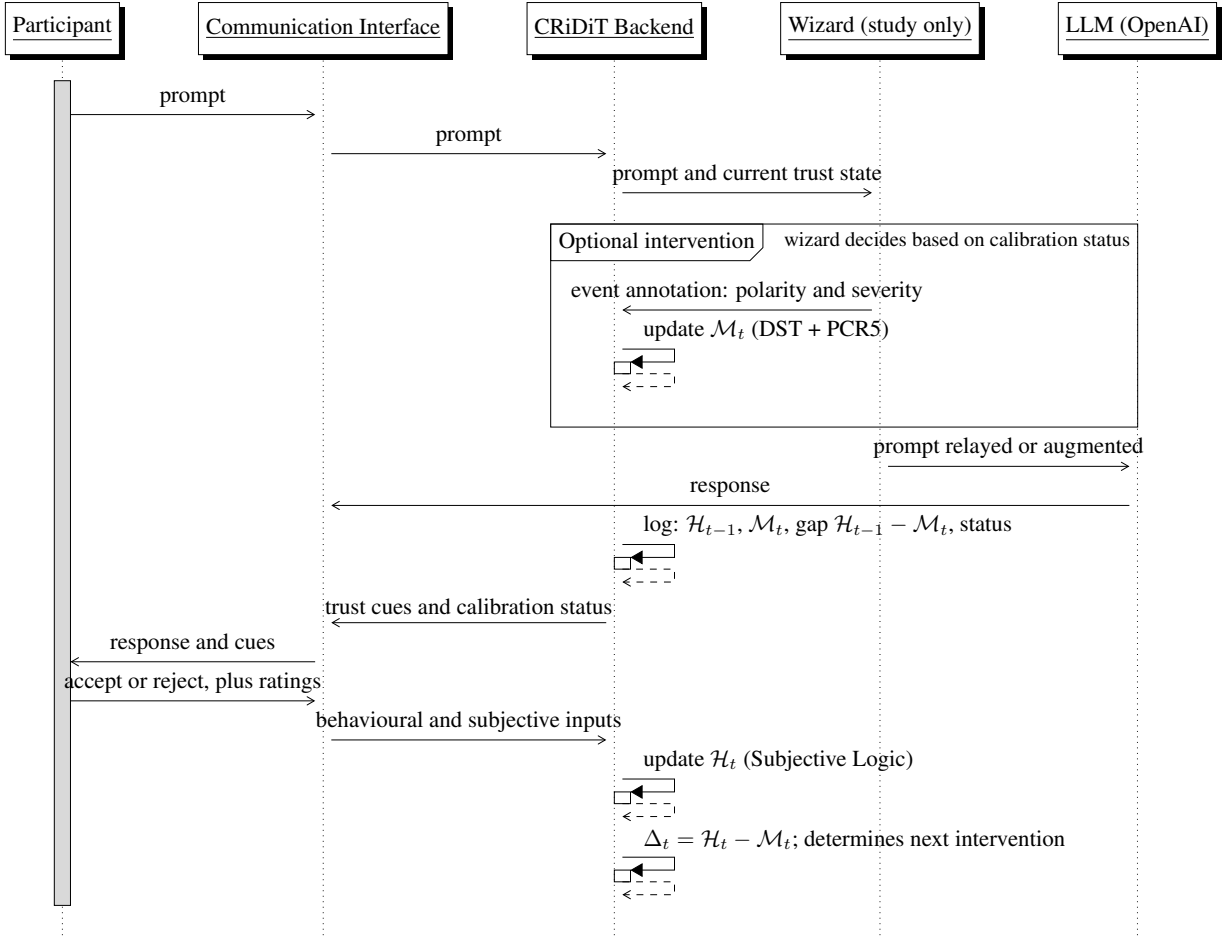

  \centering
  \resizebox{\textwidth}{!}{%
  \begin{sequencediagram}
  \newthread{p}{Participant}
  \newinst[1.2]{ui}{Communication Interface}
  \newinst[1.2]{b}{CRiDiT Backend}
  \newinst[1.2]{w}{Wizard (study only)}
  \newinst[1.2]{l}{LLM (OpenAI)}

  \mess{p}{prompt}{ui}
  \mess{ui}{prompt}{b}
  \mess{b}{prompt and current trust state}{w}

  \begin{sdblock}{Optional intervention}{wizard decides based on calibration status}
    \mess{w}{event annotation: polarity and severity}{b}
    \begin{callself}{b}{update $\mathcal{M}_{t}$ (DST + PCR5)}{}
    \end{callself}
  \end{sdblock}

  \mess{w}{prompt relayed or augmented}{l}
  \mess{l}{response}{ui}

  \begin{callself}{b}{log: $\mathcal{H}_{t-1}$, $\mathcal{M}_{t}$,
    gap $\mathcal{H}_{t-1} - \mathcal{M}_{t}$, status}{}
  \end{callself}
  \mess{b}{trust cues and calibration status}{ui}
  \mess{ui}{response and cues}{p}

  \mess{p}{accept or reject, plus ratings}{ui}
  \mess{ui}{behavioural and subjective inputs}{b}
  \begin{callself}{b}{update $\mathcal{H}_{t}$ (Subjective Logic)}{}
  \end{callself}
  \begin{callself}{b}{$\Delta_{t} = \mathcal{H}_{t} - \mathcal{M}_{t}$;
    determines next intervention}{}
  \end{callself}
\end{sequencediagram}}
  \caption{One interaction step of the CRiDiT loop. $\mathcal{M}$ updates at annotation; $\mathcal{H}$ updates after participant ratings. The log record at step $t$ pairs $\mathcal{H}_{t-1}$ with $\mathcal{M}_t$, so the human-side response to a machine-side event appears in the following record. The wizard lane is used only in the feasibility study.}\label{fig:cridit_sequence}
\end{figure}

Figure~\ref{fig:cridit_sequence} shows one interaction step. The pre-flight questionnaire initialises $\mathcal{H}_0$; observers set $\mathcal{M}_0$ and $\tau$. At each step, the participant submits a prompt, the observer optionally annotates an event and augments the prompt, the model responds, and the participant rates the response. CRiDiT updates $\mathcal{M}_t$ at the annotation step and $\mathcal{H}_t$ after the ratings are submitted, then recomputes $\Delta_t$ and determines whether the next step requires intervention.

\section{Observational Case Study Design}\label{sec:methodology}


Following Design Science Research, the observational study investigates how well the CRiDiT prototype addresses R1--R6 and produces interaction logs from which RQ1 and RQ2 can be answered. It is not designed to establish causal effects or generalise to user populations. Its purpose is demonstrative evaluation: showing that CRiDiT instruments trust calibration dynamics in practice, and surfacing where the instantiated policy departs from its requirements. The findings feed directly into the second set of requirements elicited in Section~\ref{sec:next_design_cycle}.

\subsection{Data Collection}
We collect data through two streams: questionnaires for subjective data and observational protocols for behavioural data.

\subsubsection{Pre- and Post-Flight Questionnaire}
The questionnaires enable the assessment of trust patterns and correlations between perceived trustworthiness and individual characteristics.

Participants completed a pre-flight questionnaire assessing baseline states and contextualising interaction expectations, covering:
\begin{itemize}
  \item dispositional trust in AI technology (including trust intention, risk aversion, and privacy concern);
  \item their familiarity with the domain;
  \item the importance assigned to trust factors relevant to the scenario;
  \item two open questions on what would break their trust and what would foster their trust in the chatbot.
\end{itemize}

After the interaction phase, participants completed a post-flight questionnaire evaluating the interaction outcome and their perception of the system, covering:
\begin{itemize}
  \item post-interaction trust in the chatbot;
  \item willingness to continue using the chatbot;
  \item perceived performance on trust factors;
  \item overall assessment of the session;
  \item two open questions on their perception of trust evolution during the interaction phase and their expectations going forward.
\end{itemize}

Except for the open questions, all items use a 1--7 Likert scale.

\subsubsection{Observational Protocol}
The observational protocol allows the collection of real-time qualitative behavioural data during the interaction. For each decision point or system output within a scenario, participants were presented with trust cues as references, and data were collected on:
\begin{itemize}
  \item CRiDiT-generated outputs: machine trustworthiness score $\mathcal{M}_t$, human trust estimate $\mathcal{H}_t$, calibration gap $\Delta_t$, calibration status, and static warnings depending on the calibration status;
  \item self-reported ratings on self-confidence, outcome reliability, outcome predictability, and task criticality (1--7 Likert scale);
  \item think-aloud feedback;
  \item behavioural choices: accepting, rejecting, or querying the system further.
\end{itemize}

\subsection{Participants and Procedure}
We recruited 15 participants, five per scenario. Each participant experienced one scenario in a between-subjects design. Within each group, three participants had relevant background in the scenario (graduates or professionals in the domain) and two had no domain-specific experience, except in the financial scenario where the split was reversed. Expert participants included graduate-level individuals with at least two years of experience in the respective domain. Non-expert participants were doctoral students without domain-specific experience. All participants were able to understand and complete the assigned tasks using the given task descriptions and artificial data.

Participants were informed that the chatbot used a general-purpose language model accessed through the OpenAI API. They received an interface tutorial, task descriptions, artificial data, and instructions for think-aloud reporting and scaled feedback. Each participant completed two or three tasks in the assigned scenario, reported whether they accepted or rejected each response, and provided self-confidence, reliability, predictability, and task criticality ratings. The interface displayed numerical estimates, calibration status, and static warnings when applicable. The study wizard's interventions followed a Wizard-of-Oz protocol~\cite{kelley1984iterative} without the participants' knowledge. After the interaction phase, participants evaluated system performance and reported changes in their trust.

\subsection{Data Processing}
We process the interaction logs to derive metrics for analysing participant responses to trust violations and remediation actions. The CRiDiT backend generates three data files: pre-flight questionnaire responses, interaction logs, and post-flight questionnaire responses. 

Pre-flight responses characterise participants' prior attitudes and expectations, including initial trust propensity and the importance they assigned to trust factors such as accuracy, explainability, and transparency. Post-flight responses capture participants' overall assessment of the system after the session.

Each interaction log identifies the session, scenario, and expertise level. For each iteration we record $\mathcal{H}_t$, $\mathcal{M}_t$, $\Delta_t$, calibration status and, when present, the prompt label (e.g., complementary explanation, uncertainty communication (explanation), or correction).

Trust factors from pre-flight questionnaires guide observer interventions: injected issues target dimensions participants consider critical. Violations lead to decreases in $\mathcal{M}_t$ and are communicated as trust indicators.
We quantify violation dynamics using:
\begin{align*}
  \Delta\mathcal{M}^v_t &= \mathcal{M}^v_t - \mathcal{M}_{t-1} \\
  \Delta\mathcal{H}^v_t &= \mathcal{H}^v_{t+1} - \mathcal{H}_t
\end{align*}

Remediation events are identified from prompt content and manual recoding of the label field, as described in Section~\ref{subsec:operationalisation}. We measure the immediate effect of remediation on the human side as:
\begin{equation*}
  \Delta\mathcal{H}^r_t = \mathcal{H}^r_{t+1} - \mathcal{H}_t
\end{equation*}
The machine-side effect is computed as $\Delta\mathcal{M}^r_t = \mathcal{M}_t - \mathcal{M}_{t-1}$, where $\mathcal{M}_t$ reflects evidence annotated by the observer at step $t$. Because $\mathcal{M}_t$ only updates when an observer explicitly annotates an event, autonomous system behaviour that a participant detects but the observer does not annotate leaves $\mathcal{M}_t$ unchanged. Human trust may nonetheless shift in response, so the resulting gap appears in the log as miscalibration rather than as an unrecorded event. Isolating these cases requires automated evidence capture, which we identify as a requirement for the next design cycle.

\subsection{Abductive Inference Design}
For an observational case study, think-aloud feedback and trust-related behaviour are difficult to capture through quantitative metrics alone. Despite the statistical analysis on quantified trust score evolution and the calibration event rate reflecting trust patterns, we need to bring further explanations to the phenomena and observations found during execution.

We therefore apply abductive inference to bring explanations to the observed patterns, especially when unexpected issues occur (e.g., task misunderstandings, unexpected participant behaviour). The analysis proceeds in three steps: (1) we collect descriptive summaries from raw interaction data (participant feedback and explanations of their behaviour); (2) we generate explanations for observed patterns in trust evolution, grounded in the CRiDiT interaction context (including prompts and remediation actions); and (3) we formulate generalisations that specify the scope in which these explanations plausibly hold.

\section{Results}\label{sec:result}
This section reports the analysis of the interaction logs produced by the study: Section~\ref{sec:reported_factors} gives the context participants described; Sections~\ref{sec:rq1} and~\ref{sec:rq2} address the research questions in turn.

\subsection{Interaction Context from Participant Reports}\label{sec:reported_factors}
Because the recorded estimates are aggregated from different trust factors, they allow us to track how trust changes but not to determine which factors participants were responding to. Post-flight questionnaire responses and think-aloud remarks provide contextual information; we treat them as supplementary evidence rather than findings about trust in general.

Across the three scenarios participants repeatedly identified accuracy-related and verifiability-related factors, particularly reasoning flow and source grounding, as the most frequently weighted. This was especially pronounced in the financial and legal tasks involving risk analysis and case-based reasoning. Communication and presentation quality also mattered: structure, clarity and tone influenced perceived reliability even where content was correct. Remarks ranged from mild preference (``Bullet lists are better than text'') to something stronger (``Inaccurate terms and vocabulary in the legal scenario reduced my trust. I know it may be due to translation problems, but for professional work, misuse of precise terms destroys trust'').

When outputs were insufficient, participants issued follow-up queries and reformulated prompts. Repeated prompting was itself read as a lack of logical fluency. Tolerance to violations varied by task: participants reported stricter requirements where errors were unacceptable (precise computation, classification, source precision) but accepted incompleteness on open-ended tasks. One participant framed the trade-off directly: ``On the one side, I want the risk analysis to be as complete as possible, but since this is a general chatbot without domain-specific RAG, I do not have the energy to read everything.''

Two observations bear on what the artifact can represent. First, participants sometimes introduced malformed prompts without the system flagging them, so degradation arising on the human side entered $\mathcal{M}$ as though it had originated in the system. Second, participants distinguished sharply between tolerating incompleteness on open-ended tasks and rejecting inaccuracy where correctness criteria were clear, a distinction that neither a threshold fixed per scenario nor an estimate fused across rating dimensions can carry. In both cases the artifact's representation is coarser than the judgement it is meant to track, which is the thread taken up in Section~\ref{sec:implications}.

\subsection{RQ1: What calibration dynamics does the artifact capture at run time?}\label{sec:rq1}
The artifact captures three dynamics continuously across the 144 logged interaction steps: the divergence between $\mathcal{H}_t$ and $\mathcal{M}_t$ as each estimate responds to its respective inputs, the largely absent human-side response to machine-side degradation, and the type-dependent effect of interventions on each side of the loop. Section~\ref{sec:traced_calibration} traces one session in detail to make these concrete; Sections~\ref{sec:states} through~\ref{sec:remediation} report the corresponding patterns across the full interaction logs.

\subsubsection{Traced Session}\label{sec:traced_calibration}
To make the artifact's run-time behaviour inspectable, we trace one complete session step by step. We select a non-expert participant in the legal scenario ($\tau = 0.06$), the condition in which under-trust states are most frequent (12 of 18 steps, Table~\ref{tab:states}). Figure~\ref{fig:legal_trust_evolution} shows the full trajectory; the remaining fourteen sessions appear in Appendix~\ref{app:traces}.

\begin{figure}[h]
  \centering
  \includegraphics[width=0.7\textwidth, height=0.29\textheight]{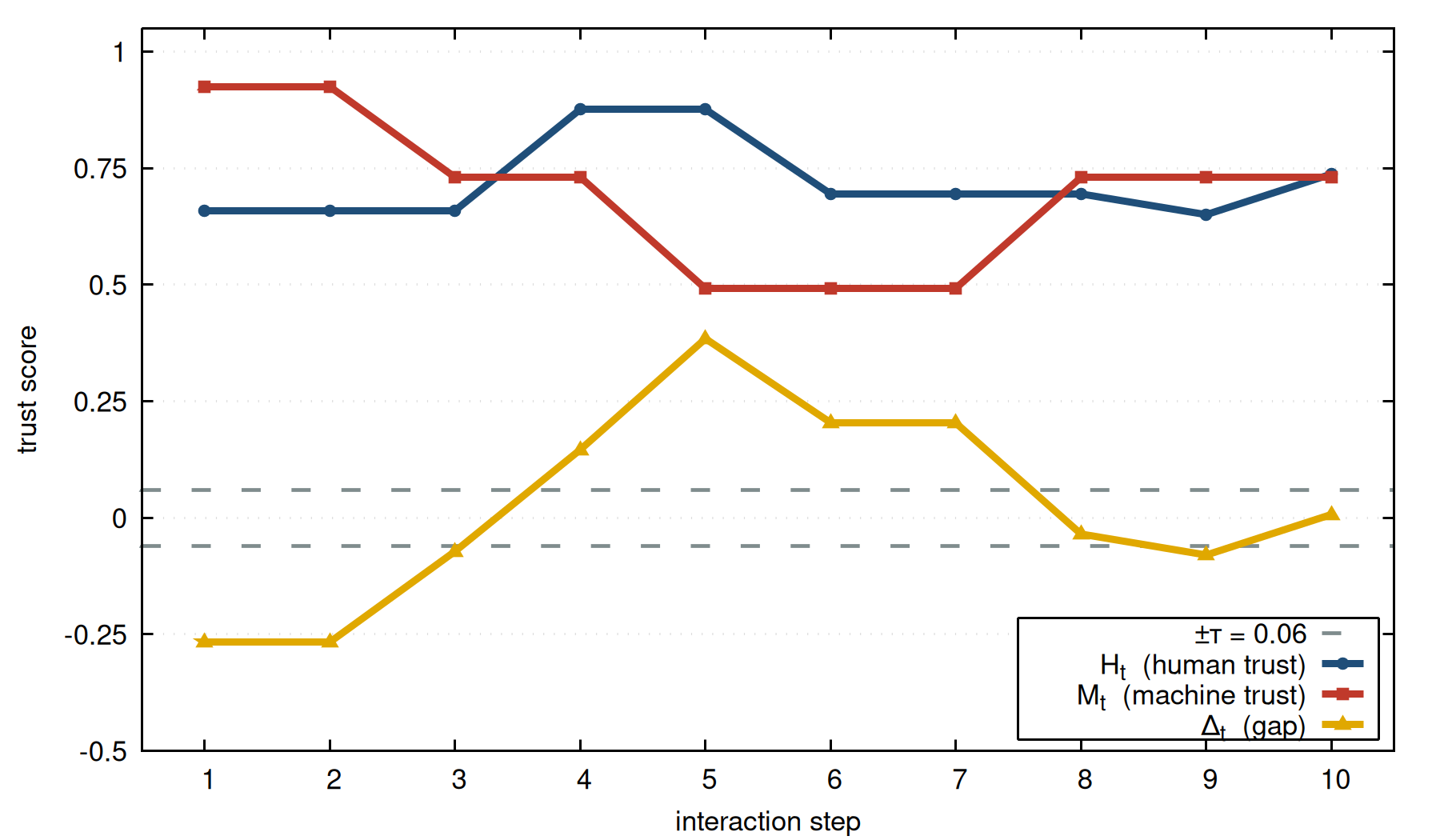}
  \caption{Trust evolution for a legal non-expert participant.}\label{fig:legal_trust_evolution}
\end{figure}

At $t_1$ the pre-flight questionnaire initialises $\mathcal{H}_0 = 0.658$ against a benchmark-derived $\mathcal{M}_0 = 0.925$, placing the participant in under-trust ($\Delta = -0.267$) before any interaction has occurred. From $t_2$ onward each step updates $\mathcal{H}$ from the participant's self-reported rating and adoption decision, and $\mathcal{M}$ from wizard-annotated response events.

A first degradation at $t_3$ lowers $\mathcal{M}$ to $0.730$. Human trust does not move; because the participant began in under-trust, the gap merely narrows to $-0.072$ and no remediation is triggered. At $t_4$ self-reported trust rises to $0.876$ with no corresponding change in machine trust, and the gap crosses into over-trust ($+0.146$). A second degradation at $t_5$ lowers $\mathcal{M}$ to 0.492 while $\mathcal{H}$ remains unchanged, widening the gap to $+0.384$, the largest in this session. The participant's response arrives at $t_6$: $\mathcal{H}$ falls to 0.650 and the gap narrows to $+0.158$. This one-step offset follows from the update rule, since $\mathcal{M}$ is revised when the observer annotates an event, before the response reaches the participant, and $\mathcal{H}$ only after the participant has rated that response, so a degradation and its perception are necessarily recorded at consecutive steps. The over-trust state logged at $t_5$ is therefore transient by construction rather than a period of sustained miscalibration.

A corrective prompt at $t_8$ raises $\mathcal{M}$ back to $0.730$, closing the gap to $-0.035$, while $\mathcal{H}$ remains unchanged. The participant's response arrives at $t_9$: $\mathcal{H}$ falls to $0.650$ ($\Delta\mathcal{H} = -0.045$), widening the gap to $-0.080$, as the participant flagged an inconsistency in the corrected response. An explanatory prompt at $t_9$ accompanies the response; $\mathcal{H}$ recovers to $0.737$ at $t_{10}$ and the gap closes to $+0.007$, within $\tau$, after which no further remediation is triggered.

Two properties of the closed loop are visible here. First, the artifact recorded a non-expert participant detecting an injected degradation, with the detection appearing one step after the evidence entered on the machine side. Second, remediation was triggered on 7 of 10 steps, a rate driven by the wide initial gap that $\mathcal{M}_0$ imposes; we examine this configuration effect in Section~\ref{sec:baseline}.

\subsubsection{Calibration States}\label{sec:states}
Table~\ref{tab:states} gives the calibration state recorded at each step, grouped by scenario and participant expertise. The distribution is uneven across conditions: financial non-experts recorded over-trust on 20 of 27 steps and no well-calibrated steps at all, while hiring experts recorded over-trust on 15 of 34. The legal condition is characterised by under-trust for both groups (11 of 32 expert steps, 12 of 18 non-expert steps). These proportions should be read against the baseline effect established in Section~\ref{sec:baseline}, since every session begins in under-trust by construction.

\begin{table}[h]
  \caption{Calibration states across all 144 logged steps, recomputed under the intended per-scenario thresholds ($\tau = 0.06$ legal, $0.07$ financial, $0.08$ hiring). Every session opens in under-trust by construction.}
  \label{tab:states}
  \begin{tabular}{llrrrr}
    \toprule
    Scenario & Expertise & Steps & Over-trust & Well-calibrated & Under-trust \\
    \midrule
    Hiring    & Expert     &  34 & 15 & 11 & 8 \\
    Hiring    & Non-Expert &  15 & 5  & 4  & 6 \\
    Financial & Expert     &  18 & 6  & 8  & 4 \\
    Financial & Non-Expert &  27 & 20 & 0  & 7 \\
    Legal     & Expert     &  32 & 5  & 16 & 11 \\
    Legal     & Non-Expert &  18 & 4  & 2  & 12 \\
    \bottomrule
  \end{tabular}
\end{table}

\subsubsection{Human Response to Machine-side Degradation}\label{sec:degradation}
Across 25 degradation events, $\mathcal{H}$ did not change at all in 16 cases (median $\Delta\mathcal{H}^v_t = 0.000$), and this holds for experts and non-experts alike. Degradation was thus more often absorbed than registered. Where $\mathcal{H}$ did move, the direction was not consistent: hiring experts showed a mean decrease of $-0.068$ across 7 events (5 of which were unchanged and 2 negative), while financial non-experts showed a mean increase of $+0.053$ across 6 events. The one-step offset established in Section~\ref{sec:traced_calibration} applies: $\Delta\mathcal{H}^v_t = \mathcal{H}^v_{t+1} - \mathcal{H}_t$ captures the participant's response in the step following the degradation.

\subsubsection{Response to Remediation}\label{sec:remediation}
Table~\ref{tab:interventions} shows the effect of each intervention type across all steps. The three intervention types were associated with different changes on the two sides of the loop. All 9 corrections raised $\mathcal{M}$, since correcting a response constitutes positive evidence entering the machine-side estimate via the DST pipeline. Complementary explanations did not raise $\mathcal{M}$, since adding context without altering correctness generates no new evidence. Uncertainty communication raised $\mathcal{M}$, since making limits explicit is annotated as a trustworthiness signal. On the human side, explanations raised $\mathcal{H}$ more often than not (complementary: 7 of 10, uncertainty communication: 2 of 2). Corrections are more variable: $\mathcal{H}$ rose on 5 of 9 steps and fell on 3, suggesting that a correction can lower perceived trust when it reveals a flaw the participant had not noticed.

\begin{table}[h]
  \caption{Effect of each intervention type across all calibration states. $\Delta\mathcal{M}_t = \mathcal{M}_{t+1} - \mathcal{M}_t$ and $\Delta\mathcal{H}_t = \mathcal{H}_{t+1} - \mathcal{H}_t$.}\label{tab:interventions}
  \begin{tabular}{lrrrr}
    \toprule
    Type & $n$ & $\Delta\mathcal{M}_t > 0$ & Mean $\Delta\mathcal{H}_t$ &
      $\Delta\mathcal{H}_t > 0$ \\
    \midrule
    Violation                               & 24 & 0 & $-0.007$ & 3 \\
    Correction                              & 9  & 9 & $+0.034$ & 5 \\
    Explanation (complementary)             & 10 & 0 & $+0.046$ & 7 \\
    Explanation (uncertainty communication) & 2  & 2 & $+0.130$ & 2 \\
    \bottomrule
  \end{tabular}
\end{table}

Table~\ref{tab:channels} breaks this down by calibration state. Under over-trust the distinction is sharpest: corrections narrowed the gap on all 6 steps, since raising $\mathcal{M}$ reduces the gap directly even when $\mathcal{H}$ moves in the same direction. Complementary explanations narrowed it on only 1 of 4 over-trust steps despite raising $\mathcal{H}$ on 3 of 4, because $\mathcal{M}$ did not move. The calibration policy specifies explanatory prompts for over-trust; the interaction logs show complementary explanations were the type issued, the type whose machine-side effect is zero by construction.

\begin{table}[h]
  \caption{Effect of each intervention type on machine trust $\Delta\mathcal{M}_t = \mathcal{M}_{t+1} - \mathcal{M}_t$ and human trust $\Delta\mathcal{H}_t = \mathcal{H}_{t+1} - \mathcal{H}_t$, by calibration state at $t-1$.}
  \label{tab:channels}
    \begin{tabular}{llrrrr}
    \toprule
    Type & State & $n$ & $\Delta\mathcal{M}>0$ & $\Delta\mathcal{H}>0$ & Gap narrowed \\
    \midrule
    Correction & Over-trust  & 6 & 6 & 3 & \textbf{6} \\
    & Under-trust & 3 & 3 & 2 & 2 \\
    \addlinespace
    Explanation (complementary) & Over-trust  & 4 & 0 & 3 & 1 \\
    & Under-trust & 1 & 0 & 1 & 0 \\
    & Well-calib. & 5 & 0 & 3 & 0 \\
    \addlinespace
    Explanation (uncertainty communication) & Over-trust  & 1 & 1 & 1 & 1 \\
    & Well-calib. & 1 & 0 & 1 & 1 \\
    \bottomrule
  \end{tabular}
\end{table}

The static warning is largely inert. On its own, it left $\mathcal{H}$ unchanged on 21 of 26 over-trust steps and on similar proportions elsewhere (26 of 38 under-trust, 19 of 29 well-calibrated): the mechanism present on every detected step produces almost no observable effect. When a prompt accompanied the warning, $\mathcal{H}$ moved on 19 of 21 such steps, but not consistently in one direction: across all states it rose on 14 and fell on 5. Whether the prompt helps or harms depends on the response it elicits.

\subsubsection{Interpretation}\label{sec:rq1_interpretation}
Taken together these results characterise what the artifact instruments rather than what users do. It maintains both estimates continuously, computes and records every threshold crossing in the interaction logs, and logs the actions triggered in response. It does not, in this configuration, isolate the effect of those actions.

The degradation results show that the human-side estimate is largely insensitive to the events driving the machine-side estimate: participants absorbed degradation without revising their trust on 16 of 25 occasions. The remediation results show that intervention type determines which side of the loop is affected: corrections enter positive evidence and raise $\mathcal{M}$, complementary explanations do not, so the two are not interchangeable in an over-trust state. The static warning, present on every detected step, produced almost no observable effect on its own.

Two limitations bound what these results can establish. First, prompt issuance was wizard-selected rather than randomised, so the contrast between intervened and non-intervened steps cannot be attributed to the intervention alone. Second, the machine-side effect of a remediation is computed from the log as $\Delta\mathcal{M}^r_t = \mathcal{M}_{t+1} - \mathcal{M}_t$, but because multiple annotations can occur at the same step, this value reflects the net evidence entered at that step rather than the remediation in isolation. Establishing causal effectiveness requires the automated intervention layer and the controlled designs discussed in Section~\ref{sec:future}.

\subsection{RQ2: Where does the instantiated policy deviate from its design requirements?}\label{sec:rq2}
The instantiation deviates from its requirements in three respects: the benchmark baseline fixes the initial calibration state before interaction begins, the risk-sensitive thresholds do not produce risk-sensitive triggering, and the calibration policy assigns explanatory prompts to over-trust, where the gap narrowed more consistently after corrections than after explanations in the interaction logs. Sections~\ref{sec:baseline}--\ref{sec:policy} establish each deviation from the interaction logs; Section~\ref{sec:requirements} assesses all six requirements against the full interaction logs.

\subsubsection{Baseline Configuration}\label{sec:baseline}
Participants reported a positive dispositional tendency to trust the chatbot (mean trust intention 5.021/7), yet all 15 sessions opened in under-trust. The two are different constructs: the first is an absolute self-report, the second is defined relative to $\mathcal{M}_0$. All 15 sessions were initialised with $\mathcal{M}_0 = 0.925$, the MMLU score of the underlying model, whereas $\mathcal{H}_0$ derived from the pre-flight questionnaire ranged from $0.540$ to $0.746$ (mean $0.658$), so the initial gap is negative by construction regardless of disposition and before any interaction has occurred. This is visible in every panel of Figure~\ref{fig:all_traces}.

\subsubsection{Threshold Configuration}\label{sec:threshold}
Across all 144 logged steps the median absolute gap is $0.148$, well above the largest configured threshold. Table~\ref{tab:tau_sweep} recomputes the share of steps that would trigger remediation under a range of thresholds, holding the logged $\mathcal{H}_t$ and $\mathcal{M}_t$ fixed. At the thresholds actually configured the artifact would have been in a remediation state on 103 of 144 steps (71.5\%): 69.4\% in hiring, 82.2\% in financial, 64.0\% in legal. Reaching a 50\% trigger rate would require $\tau \approx 0.15$, between two and four times the configured values.

\begin{table}[h]
  \caption{Share of interaction steps triggering remediation under varying $\tau$, recomputed from the logged trust trajectories.}
  \label{tab:tau_sweep}
  \begin{tabular}{lrrrr}
    \toprule
    $\tau$ & Hiring & Financial & Legal & All \\
    \midrule
    0.02     & 95.9\% & 100.0\% & 88.0\% & 94.4\% \\
    0.04     & 93.9\% & 100.0\% & 70.0\% & 87.5\% \\
    0.06$^*$ & 79.6\% &  95.6\% & 64.0\% & 79.2\% \\
    0.07$^*$ & 75.5\% &  82.2\% & 60.0\% & 72.2\% \\
    0.08$^*$ & 69.4\% &  77.8\% & 54.0\% & 66.7\% \\
    0.10     & 67.3\% &  77.8\% & 48.0\% & 63.9\% \\
    0.15     & 55.1\% &  57.8\% & 36.0\% & 49.3\% \\
    0.20     & 46.9\% &  40.0\% & 30.0\% & 38.9\% \\
    0.30     &  8.2\% &  11.1\% &  8.0\% &  9.0\% \\
    \bottomrule
  \end{tabular}
\end{table}

Thresholds were set lower for higher-risk scenarios so that calibration would be most sensitive in the legal condition. In practice legal shows the lowest trigger rate of the three, because the gaps observed there were smaller, and the ordering holds across the whole sweep in Table~\ref{tab:tau_sweep}. Sensitivity depends on the scale of the gaps each context produces as well as on $\tau$; fixing $\tau$ per scenario does not determine the sensitivity that results from it. Risk-sensitive triggering, as instantiated here, is therefore under-determined by $\tau$ alone.

\subsubsection{Calibration Policy}\label{sec:policy}
The calibration policy specifies explanatory prompts for over-trust and corrective prompts for under-trust. Table~\ref{tab:channels} shows that the assignment is inverted relative to what closes the gap: corrections narrowed the gap on all 6 over-trust steps by raising $\mathcal{M}$, whereas complementary explanations narrowed it on only 1 of 4 over-trust steps since $\mathcal{M}$ did not move. The policy specifies the type whose machine-side effect is zero by construction for the state where the machine-side channel matters most.

\subsubsection{Design Requirements Assessment}\label{sec:requirements}
We assess the instantiated artifact against each requirement, recording what the instantiation achieved and where it fell short.

\paragraph{R1 (Risk-Aware Inputs).} Risk-related inputs were collected in the pre-flight questionnaire and used both to initialise $\mathcal{H}_0$ and to configure the calibration thresholds, so risk enters the computation as the requirement asks. During interaction, however, participants rarely reported uncertainty, and the artifact could not distinguish degradation originating in the system from degradation originating in a participant's own incomplete or malformed prompt. Risk representation therefore remained static where the requirement intends it to be dynamic.

\paragraph{R2 (Risk-Aware Calibration).} Thresholds were derived from risk analysis and assigned per scenario, and remediation was triggered whenever $\Delta_t$ crossed them, so the mechanism operated as specified. The intended effect did not follow, as established in Section~\ref{sec:threshold}: sensitivity depends on the scale of the gaps each context produces as well as on $\tau$, and the ordering of trigger rates inverted the intended risk ordering. 

\paragraph{R3 (Trust Updating).} The calibration history records time-stamped updates of $\mathcal{H}_t$, $\mathcal{M}_t$, $\Delta_t$ and the resulting decision at every interaction step, which is what allowed the interaction logs to be reconstructed session by session. The threshold field in the calibration log does not reliably record the value in force, one session carries zero throughout, three carry another scenario's value; so calibration states are recomputed from the logged $\mathcal{H}_t$ and $\mathcal{M}_t$ under the intended per-scenario policy rather than read from the logged decisions. The deeper limitation is not in the updating but in what the record separates: an intervention and the estimate it revises are written at the same moment, so the effect of an intervention cannot be recovered from the log independently of the intervention itself. 

\paragraph{R4 (Human-Machine Trust Alignment).} Subjective trust signals from the questionnaires and from in-session ratings are fused into $\mathcal{H}_t$ through Subjective Logic and expressed on the same $[0,1]$ scale as $\mathcal{M}_t$, so the requirement is satisfied as written. Both quantities are interpreted as expected probabilities of trustworthiness, which is consistent with the Pignistic transform and the Subjective Logic projection, but the mapping from Likert responses to opinion parameters remains heuristic and the comparability of the resulting scalars has not been independently established. The requirement is therefore satisfied in form, and satisfying it is itself implicated in the deviations discussed in Section~\ref{sec:implications}. 

\paragraph{R5 (Machine Trust Estimation).} Technical evidence is represented in $\mathcal{M}_t$ through the DST/PCR5 aggregation pipeline, which updates the estimate from run-time events as required. Two limitations remain. The evidence-to-mass mapping uses heuristic parameters not independently validated, as disclosed in Section~\ref{subsec:operationalisation}. And the benchmark baseline from which $\mathcal{M}_t$ departs is a proxy for general capability rather than for trustworthiness on the task at hand, which placed all 15 sessions in under-trust before any exchange occurred. 

\paragraph{R6 (Calibration).} Calibration status is mapped to concrete actions and these execute at run time. The static warning left $\mathcal{H}$ unchanged on 21 of 26 over-trust steps, 26 of 38 under-trust steps. The accompanying prompts move $\mathcal{H}$, but the effect depends on type: all 9 corrections raise $\mathcal{M}$, and all 6 over-trust corrections narrowed the gap. Complementary explanations narrowed it on only 1 of 4 over-trust steps since $\mathcal{M}$ did not move. The calibration policy specifies explanatory prompts for over-trust; the interaction logs show that complementary explanations were the type issued, the type whose machine-side effect is zero by construction. 

\section{Discussion and Future Work}\label{sec:future}
\subsection{Implications for Closed-Loop Calibration Design}\label{sec:implications}
We identify three deviations from Section~\ref{sec:rq2} that are specific to this instantiation, but they are not unrelated. The first two deviations arise from the same design decision. Calibration acts on the difference between two separately estimated scalars. Since the conceptual frameworks leave this decision to the implementer, it carries more of the system's behaviour than its apparent simplicity suggests. The third arises one step later, in what the system does once the criterion has fired, and shows that the action set inherited from the surrounding literature does not cover the states the criterion is able to detect.

The first deviation concerns the initialisation. An estimator requires a starting value before situational evidence exists, and benchmark performance is the obvious candidate. The problem is not only that a benchmark measures general capability rather than capability on the task at hand, as Lee and See~\cite{lee2004trust} define appropriate reliance: it is that a benchmark score is a single number: the individual trust factors that subsequent evidence would revise cannot be recovered from it. When a participant then reports that reasoning flow is sound but source grounding is weak, the system has no decomposed baseline to update selectively; it can only shift the aggregate. In these interaction logs every session opened in under-trust before the first exchange, so the artifact reported a calibration state it had itself imposed, and that state reflected a measurement mismatch rather than a genuine assessment of the task context. Maturity models such as TCMM~\cite{steinmetz2025trust} specify how trustworthiness should be communicated, but they do not specify how it should be initialised.

The second deviation concerns the scale on which the criterion is read. Expressing risk sensitivity as a smaller threshold presumes that a gap of a given magnitude carries the same meaning in each context. This presumption does not hold, because the same threshold is strict or permissive according to the scale of the gaps that the estimators produce in that context, and the ordering intended to encode risk was in fact inverted in the observed behaviour. A difference of two quantities inherits the scales of both, and neither scale is anchored to anything outside the estimator that produced it.

The third deviation concerns what the criterion can act through. System trustworthiness is evaluated through a wide range of performance and risk metrics, whereas human trust perception relies on a narrower set of signals whose validity as proxies remains open. When one side rests on objective metrics and the other on subjective proxies of uncertain reliability, the observed gap may reflect measurement incompatibility as much as actual misalignment. The same asymmetry appears in the action set. Explanation and correction are the instruments calibration designs inherit from trust repair~\cite{de2018automation}, but they are not interchangeable: correction acts on the machine side, and explanation acts on the human side. The over-trust branch needs the former. The calibration policy specifies the latter.

What the first two have in common is that the criterion discards provenance. Reducing each side to a scalar makes the two sides comparable but indistinguishable: once $\mathcal{H}_t$ and $\mathcal{M}_t$ are numbers, nothing in $\Delta_t$ records what produced either of them. A system that cannot recover what produced a discrepancy cannot determine where it should be addressed. This suggests that a calibration criterion should operate on the evidence rather than on the scalars derived from it. The third deviation bears on the same point from the other end: an artifact that could attribute a discrepancy to one side would still need an action capable of moving that side in the required direction, and for over-trust the present design has none. Parasuraman and Riley~\cite{parasuraman1997humans} distinguish misuse from disuse as separate failure modes, but the remediation vocabulary developed since has been weighted toward disuse, and interventions that reduce reliance have been developed largely outside the trust literature. Which the over-trust branch should prefer is a question the present study cannot answer: establishing it requires distinguishing the effect of an explanation from that of a correction in the interaction logs, and that this is the obstacle rather than an insufficient number of events is itself part of what the instantiation shows. We do not establish this here. It is a design consequence of what the instantiation revealed, and the subject of the next cycle (Section~\ref{sec:next_design_cycle}).

\subsection{Next Design Cycle}\label{sec:next_design_cycle}
The deviations in Section~\ref{sec:rq2} return the cycle to problem investigation. Each identifies something the deductive requirements did not determine, and each yields a requirement for the next instantiation. Table~\ref{tab:next_requirements} states them against the requirements they revise.

\begin{table}[h]
  \caption{Requirements elicited from the instantiation. R1--R6 were derived from design goals before the artifact was built; NR1--NR5 were derived from where the running artifact departed from them.}\label{tab:next_requirements}
  \begin{tabular}{p{0.04\columnwidth} p{0.4\columnwidth} p{0.4\columnwidth} p{0.06\columnwidth}}
    \toprule
    & Observed in the instantiation & Requirement for the next cycle & Revises \\
    \midrule
    NR1 & Risk was collected before interaction only; autonomous system behaviour that participants detected was not captured in $\mathcal{M}_t$ unless the observer annotated it, so changes in $\mathcal{H}_t$ without a corresponding $\Delta\mathcal{M}_t$ may reflect unrecorded events rather than miscalibration. & CRiDiT shall monitor system behaviour continuously and capture evidence autonomously, distinguishing performance degradation from observer-injected events and from degradation originating in participant input. & R1 \\
    \addlinespace
    NR2 & Thresholds set lower for higher-risk scenarios produced the lowest trigger rate in the highest-risk condition; remediation obtained on 103 of 144 steps. & CRiDiT shall normalise triggering to the observed gap distribution or to the uncertainty carried by the estimates, rather than to a threshold fixed per scenario. & R2 \\
    \addlinespace
    NR3 & The gap reports that the two estimates disagree but not which side is responsible; interventions are recorded at the same moment as the estimate they revise, and their type is not recorded independently of the prompt text. & CRiDiT shall preserve the provenance of each side in its calibration criterion, shall timestamp interventions separately from the responses they produce, and shall record the type of each intervention as an explicit field. & R3, R4 \\
    \addlinespace
    NR4 & $\mathcal{M}_0$ taken from benchmark performance placed all 15 sessions in under-trust before the first exchange. & CRiDiT shall derive the initial machine trustworthiness estimate from evidence about performance on the task class at hand, or leave the calibration state undefined until sufficient evidence has accumulated. & R5 \\
    \addlinespace
    NR5 & The calibration policy specifies explanatory prompts for over-trust, but corrections narrowed the gap on all 6 over-trust steps while complementary explanations narrowed it on only 1 of 4. & CRiDiT shall align the calibration policy with the empirically effective action for each state: corrections for over-trust, where they raise $\mathcal{M}$ and close the gap, and explanations for under-trust & R6 \\
    \bottomrule
  \end{tabular}
\end{table}

NR1, NR2, NR4 and NR5 are local corrections: each repairs a specific decision while leaving the architecture intact. NR3 is not, since preserving provenance means the criterion can no longer be the difference of two scalars, and that is a change to the computing and calibration layers rather than to their parameters. It is also the requirement the others converge on, for the reason given in Section~\ref{sec:implications}: a criterion that discards what produced each estimate cannot determine where a discrepancy should be addressed. NR3 revises R4 in particular, and therefore revises a requirement that the instantiation satisfied as written, which is the sharpest form the cycle's output takes here.

Satisfying NR3 requires more than a change of representation. Determining which side a discrepancy belongs to requires an external reference for at least one of them, and tasks that admit verifiable alternatives can supply it: where a participant selects among options of independently known quality, choosing the weaker one is attributable to the human side. The present study could not provide such a reference, since controlling for degradation required tasks simple enough that output quality was uniformly adequate except where a fault was injected, which removes the natural variation the reference depends on. The next instantiation therefore faces a design tension between controlled injection, which fixes what counts as a degradation, and naturally varying task difficulty, which is what makes attribution observable.

All five requirements depend on replacing the Wizard-of-Oz layer. At present the observer issues prompts and assigns the polarity and severity labels from which $\mathcal{M}$ is computed, so the machine-side estimate is researcher-assigned and detection accuracy cannot be assessed. Intervention timing was likewise selected: 8 of 21 remediations followed a drop in $\mathcal{H}$ within two steps, confounding any comparison between intervened and non-intervened steps. Automating evidence annotation would make $\mathcal{M}$ independently computable and measurable against known injected faults; randomising remediation type and timing would remove the selection effect and allow the polarity question to be tested directly.

Two operator choices remain open and are testable against the interaction logs without new data: whether DST with PCR5 is more sensitive than linear weighted-sum aggregation under conflicting evidence, and whether prioritising behavioural over subjective inputs yields a more stable $\mathcal{H}_t$ than equal weighting. Both bear on NR2, since the sensitivity of the criterion depends on the operators producing the estimates it compares. 

\section{Threats to Validity}\label{sec:threats}
\paragraph{Internal validity.} Wizard-driven interventions determine both the machine-side evidence and the timing of remediation, so observed associations between remediation and trust change are not independently attributable to the remediation itself. The selection effect is visible in the interaction logs: 8 of 21 remediations followed a drop in $\mathcal{H}$ within two steps, suggesting that intervention tended to follow perceived trouble rather than being issued independently of it.

\paragraph{Construct validity.} Both estimates, $\mathcal{M}_t$ and $\mathcal{H}_t$, rest on heuristic mappings, from Likert responses to Subjective Logic opinions on the human side and from annotated events to belief masses on the machine side. The two are compared on a common $[0,1]$ scale as expected probabilities of trustworthiness, which is consistent with the Pignistic transform and the Subjective Logic projection, but the comparability of the resulting scalars has not been independently validated. Trust cues were visible throughout and were not manipulated, so their influence on the ratings from which $\mathcal{H}$ is computed cannot be separated from the system behaviour those ratings were meant to report. Pre-flight items measuring privacy concern and risk aversion are not reverse-scored, so higher concern raises rather than lowers $\mathcal{H}_0$. The evidence-to-mass mapping uses heuristic parameters chosen through piloting whose effect on the reported calibration states has not been characterised, and the sensitivity analysis in Table~\ref{tab:tau_sweep} varies $\tau$ but holds the mapping fixed.

\paragraph{External and ecological validity.} Fifteen participants across three scenarios, with one to six events per condition in the remediation analysis, support characterisation of the artifact rather than inference about users. Tasks used artificial inputs and a prototype interface, so behaviour may differ from deployed conditions.

\section{Conclusion}\label{sec:conclusion}
This paper instantiates CRiDiT as a run-time testbed that closes the loop between trust inputs, trust computation, gap detection and remediation, and reports an analysis of the 144 interaction steps it logged across three high-stakes scenarios. The artifact maintains both estimates continuously, computes and records every threshold crossing in the interaction logs, and logs the remediation triggered in response, making calibration episodes reconstructable step by step.

The analysis identifies three deviations between the instantiated policy and the requirements it was built to satisfy. The first concerns initialisation: $\mathcal{M}_0$ is drawn from a global benchmark rather than from task-relevant evidence, so the criterion begins in a state the context has not yet informed. The second concerns triggering: a threshold fixed per scenario cannot encode risk sensitivity, because sensitivity depends on the scale of the gaps the local context produces. The third concerns remediation: the effect of trust cues and intervention types on calibration has not been established, so the policy assigns actions whose consequences in each state remain unknown. All three point to the same absence: the design treats a scalar difference
as sufficient to act on, but context determines what that difference means and what can move it. None of this is visible in the conceptual model. It becomes visible only in a running system, which is the sense in which the artifact is diagnostic rather than merely demonstrative.

These findings characterise the artifact rather than its users, under the limitations set out in Section~\ref{sec:threats}. Addressing them requires the automated intervention layer described in Section~\ref{sec:next_design_cycle}, which would make $\mathcal{M}$ measurable against known injected faults and permit remediation type and timing to be randomised. That is the next design cycle, and the prerequisite for testing the causal claims this study cannot establish.

\bibliographystyle{unsrtnat}
\bibliography{bibliography} 

\newpage
\appendix

\section{Human Trust Estimation Details}\label{app:human_trust}
\subsection{Initialisation of $\mathcal{H}_0$}
The initial human trust score is the average of rescaled Likert responses $(v-1)/6$ per item, across 11 general items and the scenario-specific domain items. Fuzzy membership functions over the resulting score assign a base rate ($a^T \in \{0.35, 0.55, 0.75\}$), uncertainty ($u_0^T \in \{0.5, 0.3, 0.15\}$), and threshold for low, medium, and high membership respectively. Belief and disbelief are then: 
\begin{align*}
  b_0^T &= \max\!\left(0,\; \min\!\left(1 - u_0^T,\;
           \mathcal{H}_0 - a^T \cdot u_0^T\right)\right) \\
  d_0^T &= 1 - b_0^T - u_0^T
\end{align*}
Note that privacy concern and risk aversion items are not reverse-scored, so higher concern raises rather than lowers $\mathcal{H}_0$.

\subsection{Behavioural Opinion}
Behavioural inputs are binary (accepted or rejected). Let $r_t^{T,\text{beh}}$ be the cumulative acceptance count and $s_t^{T,\text{beh}}$ the cumulative rejection count up to step $t$, and let $W^{\text{beh}} = 2$ be the binomial prior weight and $a^{T,\text{beh}}$ the base rate. The Beta distribution over the probability of acceptance is:
\[
  \text{Beta}\!\left(p_t^{T,\text{beh}},\, \alpha,\, \beta\right)
  = \frac{\Gamma(\alpha+\beta)}{\Gamma(\alpha)\,\Gamma(\beta)}
    \left(p_t^{T,\text{beh}}\right)^{\alpha-1}
    \left(1-p_t^{T,\text{beh}}\right)^{\beta-1}
\]
where
\[
  \alpha = r_t^{T,\text{beh}} + a^{T,\text{beh}} W^{\text{beh}}, \qquad
  \beta  = s_t^{T,\text{beh}} + \left(1-a^{T,\text{beh}}\right) W^{\text{beh}}
\]
The corresponding Subjective Logic opinion on the proposition that the system is trustworthy is:
\[
  \omega_t^{\text{beh}} =
  \left(b_t^{T,\text{beh}},\; d_t^{T,\text{beh}},\; u_t^{T,\text{beh}},\;
  a^{T,\text{beh}}\right)
\]
where
\begin{align*}
  b_t^{T,\text{beh}} &=
    \frac{r_t^{T,\text{beh}}}{W^{\text{beh}} + r_t^{T,\text{beh}} + s_t^{T,\text{beh}}} \\
  d_t^{T,\text{beh}} &=
    \frac{s_t^{T,\text{beh}}}{W^{\text{beh}} + r_t^{T,\text{beh}} + s_t^{T,\text{beh}}} \\
  u_t^{T,\text{beh}} &=
    \frac{W^{\text{beh}}}{W^{\text{beh}} + r_t^{T,\text{beh}} + s_t^{T,\text{beh}}}
\end{align*}

\subsection{Subjective Opinion}
Self-reported ratings on reliability (\textit{rel}), predictability (\textit{pred}), self-confidence (\textit{selfConf}), and task criticality (\textit{criticality}) are mapped to a Subjective Logic opinion via a custom scalar mapping:
\begin{align*}
  \ell &= 0.5 + 0.2 \cdot \frac{(\text{rel}-4)+(\text{pred}-4)}{6} \\
  c    &= \frac{\text{selfConf}-1}{6} \cdot
          \left(1 - 0.7 \cdot \frac{|\text{rel}-\text{pred}|}{6}\right) \\
  a_f  &= 0.5 - 0.3 \cdot \frac{\text{criticality}-4}{6}
\end{align*}
with $u = 1-c$, $b = \ell - a_f \cdot u$, $d = 1-u-b$, clipped to $[0,1]$. The resulting opinion is $\omega_t^{\text{subj}} = (b, d, u, a_f)$.

\subsection{Weighted Fusion}
Raw weights $w^{\text{beh}} = 0.7$ and $w^{\text{subj}} = 0.3$ are transformed with $\gamma = 2$ and normalised:
\[
  \tilde{w}^{\text{beh}} =
    \frac{(w^{\text{beh}})^\gamma}
         {(w^{\text{beh}})^\gamma + (w^{\text{subj}})^\gamma},
  \qquad
  \tilde{w}^{\text{subj}} =
    \frac{(w^{\text{subj}})^\gamma}
         {(w^{\text{beh}})^\gamma + (w^{\text{subj}})^\gamma}
\]
giving effective weights $0.845$ and $0.155$. Each opinion is discounted by its weight $\tilde{w}$:
\[
  b' = \tilde{w} \cdot b, \qquad
  d' = \tilde{w} \cdot d, \qquad
  u' = 1 - \tilde{w} + \tilde{w} \cdot u
\]
The two discounted opinions $\omega_t^{\text{beh}'}$ and $\omega_t^{\text{subj}'}$ are then combined with the consensus fusion operator $\oplus$~\cite{josang2016subjective}:
\begin{align*}
  k &= u_1 + u_2 - u_1 \cdot u_2 \\
  b &= \frac{b_1 \cdot u_2 + b_2 \cdot u_1}{k} \\
  u &= \frac{u_1 \cdot u_2}{k} \\
  a &= \frac{a_1 \cdot u_2 + a_2 \cdot u_1}{k}
\end{align*}
The base rate formula omits the $-(a_1+a_2) \cdot u_1 \cdot u_2$ correction term from J{\o}sang's original definition; the result is bounded to $[0,1]$ by clipping. The correction term is omitted for implementation implicity; including it would require resolving the fused base rate iteratively, which the current architecture does not support. The human trust score is the expected probability of trustworthiness from the fused opinion:
\begin{equation*}
  \mathcal{H}_t = E(\omega_t) = b_t^T + a^T u_t^T
\end{equation*}

\section{Per-Session Calibration Traces}\label{app:traces}
\begin{figure}[h]
  \centering
  \includegraphics[width=\textwidth, height=0.8\textheight]{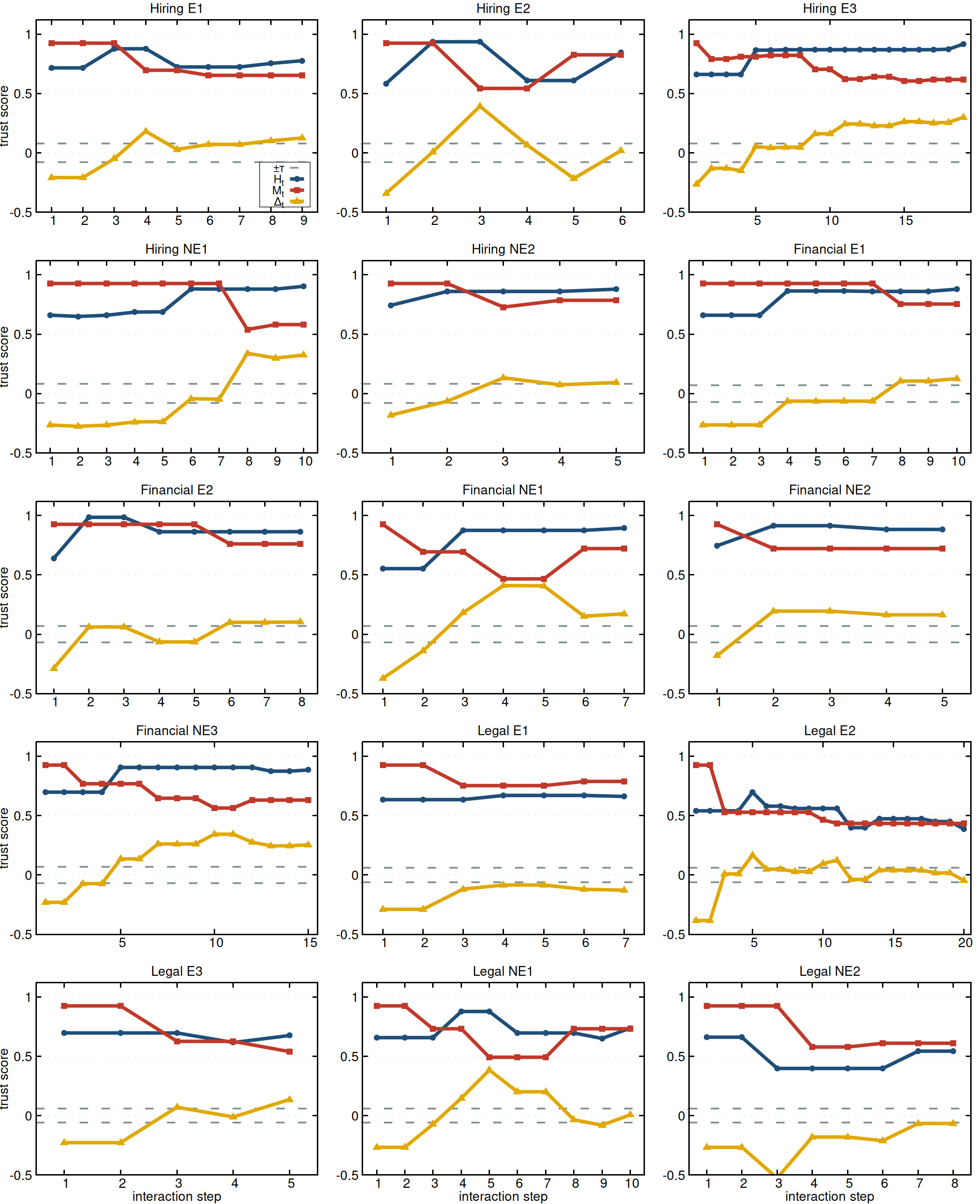}
  \caption{Run-time trust evolution for all 15 sessions. Each panel plots the human trust estimate $\mathcal{H}_t$, the machine trustworthiness estimate $\mathcal{M}_t$, and their signed gap $\Delta_t$ against interaction step, with the calibration band $\pm\tau$ shown as dashed lines at the threshold configured for that scenario. Every session opens below the lower bound, and the gap lies outside the band on 103 of the 144 steps.}\label{fig:all_traces}
\end{figure}

Figure~\ref{fig:all_traces} shows the full trajectory of every logged session. Panels are grouped by scenario and by whether the participant had relevant domain background, and share a common vertical range so that gap magnitudes are comparable across conditions.

\end{document}